\documentclass{article} % For LaTeX2e
\PassOptionsToPackage{dvipsnames,table,xcdraw}{xcolor}
\usepackage{xcolor}
\usepackage{iclr2025_conference,times}
\usepackage{graphicx}
\usepackage{math_commands}

\usepackage[utf8]{inputenc} % allow utf-8 input
\usepackage[T1]{fontenc}    % use 8-bit T1 fonts
\usepackage{hyperref}       % hyperlinks
\usepackage{url}            % simple URL typesetting
\usepackage{booktabs}       % professional-quality tables
\usepackage{amsfonts}       % blackboard math symbols
\usepackage{nicefrac}       % compact symbols for 1/2, etc.
\usepackage{microtype}      % microtypography
\usepackage{multirow}
\usepackage{algorithm}
\usepackage{algpseudocode}
\usepackage{subcaption}
\usepackage{caption}
\usepackage{wrapfig}
\usepackage{ragged2e}
\usepackage{colortbl}
\usepackage{listings}
\usepackage{fontawesome}
\usepackage{array} % Required for creating new column types

\newcolumntype{C}[1]{>{\centering\arraybackslash}p{#1}}

\makeatletter

\newcommand{\Rmnum}[1]{\expandafter\@slowromancap\romannumeral #1@}
\makeatother

\newcommand{\modelname}{AutoDAN-Reasoning}

\title{\modelname: Enhancing Strategies Exploration based Jailbreak Attacks with Test-Time Scaling}

\author{
\textbf{Xiaogeng Liu} $^{1}$ 
\;
\textbf{Chaowei Xiao}  $^{1}$   
\\
$^{1}$ Johns Hopkins University
}

\iclrfinalcopy
\begin{document}

\maketitle
\renewcommand{\thefootnote}{\fnsymbol{footnote}}
\footnotetext[1]{Corresponding to xliu316@jhu.edu}
\renewcommand{\thefootnote}{\arabic{footnote}}

\begin{abstract}
  Recent advancements in jailbreaking large language models (LLMs), such as AutoDAN-Turbo, have demonstrated the power of automated strategy discovery. AutoDAN-Turbo employs a lifelong learning agent to build a rich library of attack strategies from scratch. While highly effective, its test-time generation process involves sampling a strategy and generating a single corresponding attack prompt, which may not fully exploit the potential of the learned strategy library. In this paper, we propose to further improve the attack performance of AutoDAN-Turbo through test-time scaling. We introduce two distinct scaling methods: Best-of-N and Beam Search. The Best-of-N method generates N candidate attack prompts from a sampled strategy and selects the most effective one based on a scorer model. The Beam Search method conducts a more exhaustive search by exploring combinations of strategies from the library to discover more potent and synergistic attack vectors. According to the experiments, the proposed methods significantly boost performance, with Beam Search increasing the attack success rate by up to 15.6 percentage points on Llama-3.1-70B-Instruct and achieving a nearly 60\% relative improvement against the highly robust GPT-o4-mini compared to the vanilla method. Code is available at \url{https://github.com/SaFoLab-WISC/AutoDAN-Reasoning}
\end{abstract}

%%%%%%%%%%%%%%%%%%%%%%%%%%%%%%%%%%%%%%%%%%%%%%%%%%%%%%%%%%%%

\section{Introduction}\label{introduction}
\textit{Large Language Models} (LLMs) have become integral to numerous applications, but their widespread deployment necessitates robust safety measures to prevent misuse. Safety alignment aims to ensure LLMs behave responsibly, yet they remain vulnerable to jailbreak attacks, where crafted prompts bypass their safety protocols~\citep{NEURIPS2023_fd661313,zou2023universal,liu2024autodan}. Automated red-teaming tools are crucial for proactively identifying and mitigating these vulnerabilities.

The AutoDAN-Turbo framework~\citep{liu2025autodanturbo} represents a significant step forward in automated jailbreaking. It introduces a lifelong learning agent that autonomously discovers, evolves, and combines diverse attack strategies without human intervention. By constructing a dynamic strategy library from its interactions, AutoDAN-Turbo can generate effective and novel jailbreak prompts, achieving state-of-the-art performance in black-box settings.

However, the test-time procedure of AutoDAN-Turbo, while efficient, leaves room for improvement. After retrieving a set of relevant strategies from its library, the attacker LLM generates a single jailbreak prompt. This one-shot generation might not produce the optimal attack, even with good strategies. The inherent stochasticity and creativity of the attacker LLM mean that multiple generations from the same strategic input could yield prompts of varying effectiveness.

To address this, we propose \textbf{\modelname{}}, an enhanced version of our framework that leverages sophisticated test-time scaling methods to further amplify its jailbreak capabilities. \modelname{} utilizes the rich strategy library curated by the original AutoDAN-Turbo and introduces two powerful test-time search algorithms: Best-of-N and Beam Search. These methods allow for more deliberate and optimized exploitation of the learned strategies, significantly improving the probability of a successful jailbreak. The core contributions of \modelname{} are as follows:

\begin{itemize}
    \item \textbf{Best-of-N Scaling.} This method enhances the attack by generating multiple candidate prompts for a given strategy. Specifically, after retrieving a set of relevant strategies based on the target's previous response, the attacker LLM generates N distinct jailbreak prompts. Each prompt is then sent to the target model, and the resulting responses are evaluated by a scorer LLM. The prompt that yields the highest jailbreak score is selected as the optimal attack for that turn. This approach mitigates the risk of a suboptimal, one-shot generation by exploring a variety of prompt formulations.
    \item \textbf{Beam Search Scaling.} To explore the synergistic effects of combining multiple strategies, we introduce a beam search algorithm. This method begins by retrieving a broader pool of the top-K most relevant strategies from the library (where K is larger than in the original framework's retrieval mechanism). It then constructs a ``beam'' of the most promising initial strategy combinations. In an iterative process, these combinations are expanded by adding new strategies from the pool, and the new jailbreak prompts are generated and scored. By retaining only the top-W (beam width) combinations at each step, the algorithm efficiently navigates the vast search space of strategy compositions. This allows \modelname{} to dynamically discover and deploy the most potent combination of tactics tailored to the immediate conversational context, leading to a more sophisticated and powerful attack.
\end{itemize}

The proposed \modelname{} builds directly on the foundation of AutoDAN-Turbo, proposing a plug-in enhancement that boosts its final ASR without altering its core lifelong learning and strategy discovery mechanism. According to the experiments, the proposed method achieves significant gains across all tested models. Notably, our Beam Search approach boosts the attack success rate on Llama-3.1-70B-Instruct to 84.5\% and on the more challenging GPT-o4-mini to 33.7\%, representing absolute improvements of 15.6 and 12.5 percentage points, respectively, over the original AutoDAN-Turbo.

\begin{figure}[t!]
    \centering
    \includegraphics[width=0.9\linewidth]{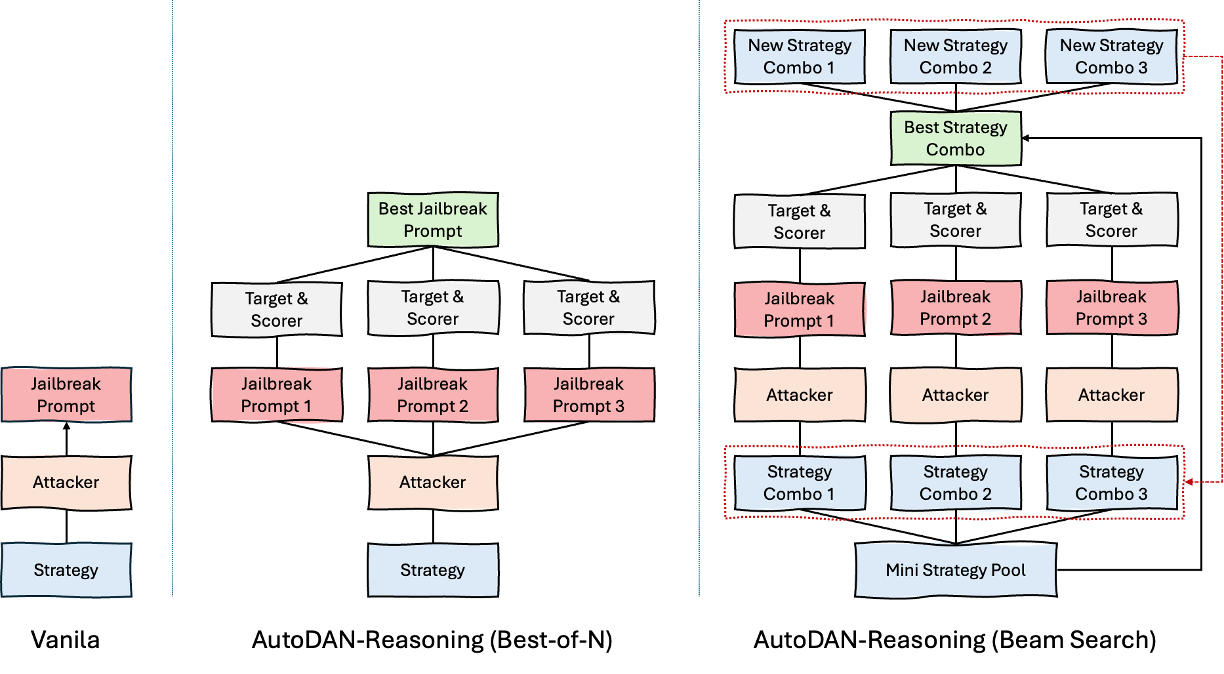}
    \caption{In this paper, we introduce Best-of-N (center) and Beam Search (right) as test-time scaling methods to improve upon the original AutoDAN-Turbo pipeline (left).}
    \vspace{-0.5cm}
    \label{fig:test_time_scaling}
\end{figure}
\section{Related Works}\label{related_works}
\textbf{Test-time Scaling.} Test-time scaling~\citep{snell2024scalingllmtesttimecompute} in large language models refers to a set of techniques that enhance the quality of a model's output by dedicating more computational resources during the inference, or ``test,'' phase. This approach improves performance without the need to retrain or alter the model's underlying parameters. Among the most prominent test-time scaling methods are Best-of-N and Beam Search. Best-of-N is a straightforward yet effective technique. It involves generating multiple (N) potential outputs from the LLM for a given input prompt. Subsequently, a separate scoring mechanism, often a reward model or a verifier trained to align with human preferences, evaluates these N candidates. The output with the highest score is then selected as the final response. Beam search is a more sophisticated algorithm that optimizes the scoring mechanism's return at each intermediate step of the solution process rather than only at the end. It maintains a ``beam'' of the most promising partial solutions and expands them step-by-step, using the scoring mechanism's per-step predictions to guide the search toward a high-quality final answer. 

\textbf{Jailbreak Attacks.} The ongoing effort to align LLMs with human values and safety principles has been met with a persistent and evolving counter-effort to circumvent these guardrails, a practice known as ``jailbreaking''~\citep{zou2023universal,chao2023jailbreaking,liu2024autodan,liu2025autodanturbo}. These attacks involve crafting malicious prompts to elicit harmful or forbidden content that the model is designed to refuse. The landscape of these attacks has matured from manual, linguistic tricks to sophisticated, automated methods that target the core computational processes of the models themselves. Early and ongoing jailbreak research has focused on prompt-level manipulations. These techniques include role-playing scenarios, where the model is asked to adopt a persona without ethical constraints~\citep{walkerspider2023Dan,shen2023do,zeng2024johnny}, and obfuscation tactics that use ciphers~\citep{yuan2024gpt4,lv2024codechameleon}, non-natural language~\citep{jiang2024artprompt}, or persuasion strategies~\citep{shen2023do} to disguise malicious intent. 

Among these attacks, AutoDAN-Turbo~\citep{liu2025autodanturbo} plays a critical and unique role by shifting the focus from crafting individual prompts or jailbreak strategies to the autonomous discovery of underlying jailbreak strategies. It operates as a lifelong learning agent, systematically exploring the attack space through a multi-agent framework. By analyzing successful and failed jailbreak attempts, it distills the core tactics into a structured ``strategy library.'' This allows AutoDAN-Turbo to not only reuse and refine effective strategies but also to combine them in novel ways, creating a dynamic and ever-evolving arsenal. This approach distinguishes it from methods that rely on predefined attack patterns or direct prompt optimization, enabling it to uncover a broader and more diverse set of vulnerabilities in a black-box setting.

A recent and significant paradigm shift in adversarial tactics has been the weaponization of test-time compute scaling—the principle of dedicating additional computational resources during inference to improve a model's performance. Originally developed to enhance model reasoning and problem-solving capabilities, these techniques have been ingeniously repurposed as powerful attack vectors. One of the leading frameworks in this domain is Adversarial Reasoning~\citep{sabbaghi2025adversarialreasoningjailbreakingtime}. This approach formulates jailbreaking as a reasoning problem, using the target model's own capabilities against it. Instead of relying on simple binary feedback (success/failure), it uses a continuous loss signal derived from the target model's output probabilities to guide a search through the prompt space. This is implemented with a system of LLMs where an ``Attacker'' proposes prompts, a ``Feedback'' module analyzes the target's response and loss to suggest improvements, and a ``Refiner'' synthesizes a new, more potent prompt. Each iteration represents a cycle of test-time compute, allowing the attack to ``think harder'' and systematically navigate toward a successful jailbreak. Another highly effective technique, LIAR~\citep{beetham2025liarleveraginginferencetime}, reframes the problem by treating the jailbreak itself as an alignment task. LIAR employs a best-of-N sampling strategy, a classic test-time compute method, but for malicious ends. An auxiliary ``prompter'' LLM generates N candidate adversarial suffixes for a harmful query. These are then evaluated against the target model, and the suffix that most successfully elicits a harmful response is selected, effectively using the target's behavior as an ``unsafe'' reward signal.

While methods like Adversarial Reasoning and LIAR demonstrate the power of test-time compute, they primarily focus on directly generating or refining attack strings. In this paper, we equip the state-of-the-art black-box jailbreak method, AutoDAN-Turbo, with a test-time scaling mechanism that operates at a higher level of abstraction (i.e., the jailbreak strategies). By integrating Best-of-N and Beam Search algorithms into its inference process, we allow AutoDAN-Turbo to more deliberately explore its vast library of learned semantic strategies, identifying not only the most potent jailbreak prompts based on retrieved strategies but also their most effective combinations in real-time. This synthesis of lifelong strategy learning and test-time optimization allows our enhanced framework, \modelname{}, to significantly further enhance attack performance and adaptability against even the most robustly aligned models.
\section{\modelname}\label{methods}
We first provide a brief overview of the original AutoDAN-Turbo framework and then detail our proposed test-time scaling enhancements.

\subsection{Recap of the AutoDAN-Turbo Framework}
The AutoDAN-Turbo pipeline consists of three main modules:
\begin{itemize}
    \vspace{-0.1cm}
    \item \textbf{Attack Generation and Exploration:} An attacker LLM generates jailbreak prompts based on strategies, a target LLM provides responses, and a scorer LLM evaluates the maliciousness of the responses.
    \vspace{-0.1cm}
    \item \textbf{Strategy Library Construction:} A summarizer LLM analyzes successful attack logs to extract, define, and store new jailbreak strategies in a library. The key for retrieval is the embedding of the refusal response that the strategy successfully overcame.
    \vspace{-0.1cm}
    \item \textbf{Jailbreak Strategy Retrieval:} During an attack, the system retrieves the most relevant strategies from the library based on the target's current response to guide the next generation attempt of the attacker LLM.
    \vspace{-0.1cm}
\end{itemize}
At test time, the attacker LLM generates an initial jailbreak prompt. Based on the target LLM's response, the system retrieves the most relevant strategies from the learned jailbreak strategy library to guide the attacker LLM in generating a new jailbreak prompt for the next iteration. This process repeats for a fixed number of iterations or until a successful jailbreak is achieved.

\subsection{Test-Time Scaling}
The core idea of our proposed test-time scaling is to redefine the prompt generation phase. Rather than using the retrieved strategies to directly generate a single jailbreak prompt, our method initiates an exhaustive search process to discover the optimal prompt that maximizes the attack perfromance.

\subsubsection{Best-of-N Scaling}
The Best-of-N method is the simpler of our two proposals. The procedure at each attack iteration during the test stage is modified as follows:
\begin{enumerate}
    \item \textbf{Strategy Retrieval:} As in the original framework, retrieve the most relevant strategies from the library based on the target's previous response.
    \item \textbf{N-Candidate Generation:} Instruct the attacker LLM to generate N distinct jailbreak prompts using the retrieved strategies.
    \item \textbf{Parallel Evaluation:} Feed all N candidate prompts to the target LLM to get N responses. Then, use the scorer LLM to assign a score to each of the N responses.
    \item \textbf{Selection and Refinement:} Select the prompt that elicited the highest-scoring response as the attack for the current iteration. This prompt-response pair and its score inform the strategy retrieval for the next iteration. To further guide the optimization process, the attacker LLM is also prompted with the this jailbreak attempt and its result.
\end{enumerate}
This method increases the probability of finding a highly effective prompt by exploring variations in the attacker LLM's output for a fixed strategy. The process is formalized in Algorithm \ref{alg:best_of_n}.

\begin{algorithm}
\caption{Best-of-N Scaling}\label{alg:best_of_n}
\begin{algorithmic}[1]
\Procedure{BestOfNAttack}{$R_{prev}, N$}
    \State $\mathcal{S} \gets \text{RetrieveStrategies}(\text{library}, R_{prev})$
    \State $\text{Candidates} \gets \emptyset$
    \For{$i \gets 1 \text{ to } N$}
        \State $P_i \gets \text{AttackerLLM.generate}(\mathcal{S})$
        \State $R_i \gets \text{TargetLLM.respond}(P_i)$
        \State $score_i \gets \text{ScorerLLM.evaluate}(R_i)$
        \State $\text{Candidates.add}((P_i, R_i, score_i))$
    \EndFor
    \State $(P_{best}, R_{best}, score_{best}) \gets \arg\max_{(P,R,s) \in \text{Candidates}} s$
    \State \textbf{return} $P_{best}, R_{best}, score_{best}$
\EndProcedure
\end{algorithmic}
\end{algorithm}

\subsubsection{Beam Search Scaling}
The Beam Search method aims to find the best combination of strategies. This is particularly useful as AutoDAN-Turbo's library contains many atomic strategies, and their combination can lead to significantly stronger attacks. We define a beam search over the space of strategy combinations.

Let the beam width be $W$ and the maximum combination size be $C$.
\begin{enumerate}
    \item \textbf{Initial Pool Selection:} At the start of the attack, retrieve a larger-than-usual pool of the top-K most relevant strategies from the library, where $K > W$.
    \item \textbf{Initialization (t=1):} For each of the top-W individual strategies from the pool, generate a prompt and get its score. These W strategies (and their corresponding prompts/scores) form the initial beam.
    \item \textbf{Expansion (t > 1):} For each of the $W$ strategy combinations currently in the beam, create new candidate combinations by appending one unused strategy from the initial pool of K strategies. This creates a set of new, larger combinations.
    \item \textbf{Evaluation and Pruning:} Generate a prompt for each new candidate combination and obtain its score. From all the newly generated candidates, select the top-W performing combinations to form the new beam for the next step.
    \item \textbf{Termination:} Repeat the expansion and pruning steps up to the maximum combination size $C$. The final output of the current turn is the prompt generated from the highest-scoring strategy combination found throughout the search. In the next turn of generation, to further guide the optimization process, the attacker LLM is also prompted with the this jailbreak attempt and its result.
\end{enumerate}
This beam search approach enables the system to systematically generate and assess intricate strategy combinations. This surpasses the original AutoDAN-Turbo's approach of using only a narrow top-K (where K is relatively small) selection of retrieved strategies, and facilitates the discovery of potent, synergistic combinations during the testing phase. The process is formalized in Algorithm \ref{alg:beam_search}.

\begin{algorithm}
\caption{Beam Search Scaling}\label{alg:beam_search}
\begin{algorithmic}[1]
\Procedure{BeamSearchAttack}{$R_{prev}, W, C, K$}
    \State $\mathcal{S}_{pool} \gets \text{RetrieveTopKStrategies}(\text{library}, R_{prev}, K)$
    \State $\text{Beam} \gets \emptyset$
    \For{$i \gets 1 \text{ to } W$}
        \State $\text{combo}_i \gets \{\mathcal{S}_{pool}[i]\}$
        \State $P_i \gets \text{AttackerLLM.generate}(\text{combo}_i)$
        \State $score_i \gets \text{ScorerLLM.evaluate}(\text{TargetLLM.respond}(P_i))$
        \State $\text{Beam.add}((\text{combo}_i, P_i, score_i))$
    \EndFor
    \For{$c \gets 2 \text{ to } C$}
        \State $\text{Candidates} \gets \emptyset$
        \For{$(\text{combo}, P, score) \in \text{Beam}$}
            \For{$\mathcal{S}_{new} \in \mathcal{S}_{pool} \setminus \text{combo}$}
                \State $\text{new\_combo} \gets \text{combo} \cup \{\mathcal{S}_{new}\}$
                \State $P_{new} \gets \text{AttackerLLM.generate}(\text{new\_combo})$
                \State $score_{new} \gets \text{ScorerLLM.evaluate}(\dots)$
                \State $\text{Candidates.add}((\text{new\_combo}, P_{new}, score_{new}))$
            \EndFor
        \EndFor
        \State $\text{Beam} \gets \text{Top W candidates from Candidates by score}$
    \EndFor
    \State $(P_{best}, R_{best}, score_{best}) \gets \arg\max_{(P,R,s) \in \text{Beam}} s$
    \State \textbf{return} $P_{best}, R_{best}, score_{best}$
\EndProcedure
\end{algorithmic}
\end{algorithm}

\section{Experiments}\label{experiments}
\subsection{Experiments Setup}\label{experiments_setup}
\textbf{Datasets}. 
We evaluate our method and baselines on the Harmbench dataset~\citep{mazeika2024harmbench}, which contains 400 diverse and malicious requests. These requests are difficult to find via search engines, making the dataset an excellent benchmark for assessing the practical risks of jailbreak attacks on LLMs.

\textbf{Large Language Models}. We evaluate our method on both open-source and closed-source large language models. The open-source targets are Llama-3.1-8B-Instruct and Llama-3.1-70B-Instruct~\citep{dubey2024llama3herdmodels}, while the closed-source target is GPT-o4-mini\footnote{Version: 2025-04-16}~\citep{openai2025o3o4mini}. For all experiments, we enforced deterministic generation by setting the temperature to zero and limited the maximum output to 4096 tokens. We use DeepSeek-R1~\cite{deepseekai2025deepseekr1incentivizingreasoningcapability} as both the attacker and scorer for our \modelname{} and the original AutoDAN-Turbo.

\textbf{Metrics}.
Following the original AutoDAN-Turbo's evaluation settings, to ensure a fair and standardized evaluation of jailbreak attacks, we employ the Harmbench Attack Success Rate (ASR), which uses a fine-tuned Llama-2-13B classifier to determine if a response is both relevant to the query and harmful~\citep{mazeika2024harmbench}. Higher values indicate more effective jailbreak methods.

\textbf{Implementation}. To evaluate the specific improvement from our test-time scaling method, we compare it against AutoDAN-Turbo using their officially provided~\footnote{\url{https://github.com/SaFoLab-WISC/AutoDAN-Turbo}}, pre-trained jailbreak strategy library. We utilize the AutoDAN-Turbo-R version to support reasoning LLMs like DeepSeek-R1~\cite{deepseekai2025deepseekr1incentivizingreasoningcapability}. Hyperparameters, such as N for the best-of-N strategy, will be detailed in the corresponding results sections.

\subsection{Main results}\label{main_results}
\begin{table}[t]
\centering
\caption{Attack success rate (\%) on different models}
\label{tab_main}
\setlength{\tabcolsep}{2pt} % Adjust column spacing if needed
\begin{small}
\begin{tabular}{r | C{2.4cm} | *{3}{C{1.1cm}} | *{3}{C{1.1cm}}}
\toprule
    & AutoDAN-Turbo & \multicolumn{3}{c|}{Turbo-Search (Best-of-N)} & \multicolumn{3}{c}{Turbo-Search (Beam Search)} \\
\cmidrule(lr){2-2} \cmidrule(lr){3-5} \cmidrule(lr){6-8} % Use trimmed rules for better grouping
    & N=1   & N=2   & N=4   & N=8   & W=2   & W=4   & W=8 \\
\midrule
Llama-3.1-8B-Ins  & 67.8  & 71.0  & 78.6  & 79.4  & 68.1  & 76.4  & 81.3 \\
Llama-3.1-70B-Ins & 68.9  & 72.4  & 81.2  & 82.0  & 70.3  & 83.6  & 84.5 \\
GPT-o4-mini            & 21.2  & 22.4  & 24.9  & 26.1  & 26.8  & 28.0  & 33.7 \\
\bottomrule
\end{tabular}
\end{small}
\end{table}
\begin{figure}[t]
    \centering 
    \begin{subfigure}{0.32\textwidth}
        \includegraphics[width=\linewidth]{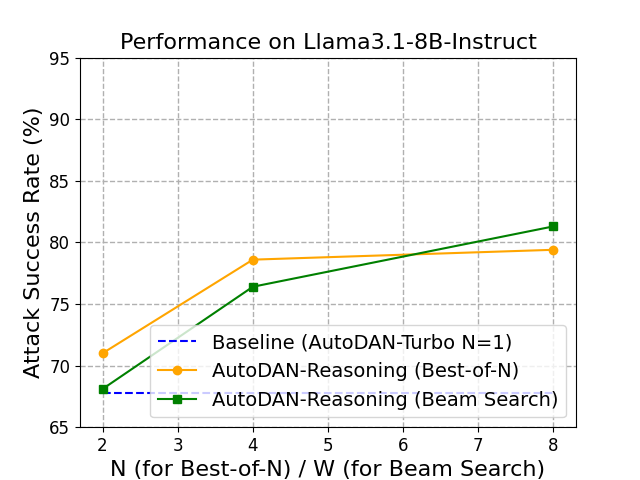}
        \caption{Llama3.1-8B-Instruct}
        \label{subfig:7b}
    \end{subfigure}
    \hfill 
    \begin{subfigure}{0.32\textwidth}
        \includegraphics[width=\linewidth]{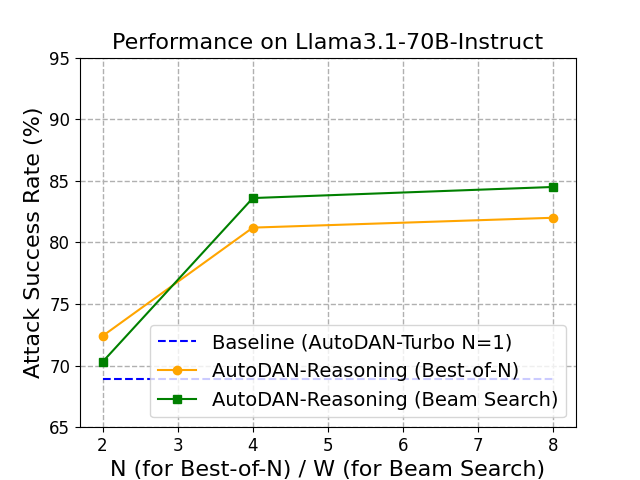}
        \caption{Llama3.1-70B-Instruct}
        \label{subfig:70b}
    \end{subfigure}
    \hfill 
    \begin{subfigure}{0.32\textwidth}
        \includegraphics[width=\linewidth]{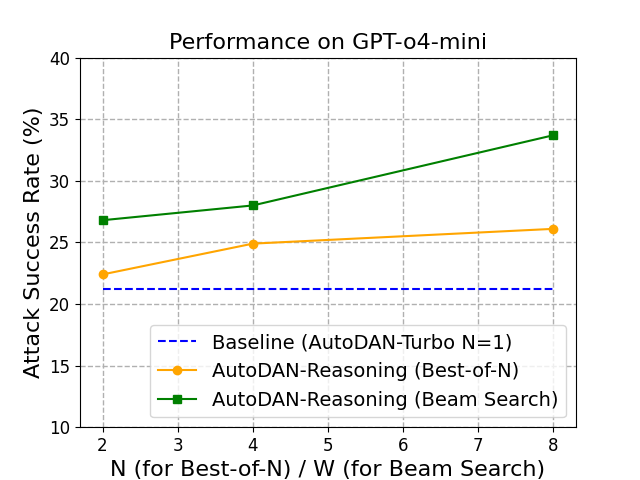}
        \caption{GPT-o4-mini}
        \label{subfig:o4}
    \end{subfigure}
    \caption{Comparison of attack success rates using \modelname{} with Best-of-N and Beam Search test-time scaling strategies across three target models. The dashed blue line represents the baseline (vanilla AutoDAN-Turbo) performance.}
    \label{fig:decoding_strategy_comparison}
\end{figure}
Our main results, presented in Tab.~\ref{tab_main} and Fig.~\ref{fig:decoding_strategy_comparison}, demonstrate that both proposed test-time scaling methods, Best-of-N and Beam Search, significantly enhance the attack performance of AutoDAN-Turbo across all tested models.

The Best-of-N method provides a consistent and notable improvement over the baseline AutoDAN-Turbo, which is equivalent to N=1. As the number of candidate prompts ($N$) increases, the Attack Success Rate (ASR) rises accordingly for every target model. On Llama-3.1-8B-Instruct, increasing $N$ from 1 to 8 boosts the ASR from 67.8\% to 79.4\%, an absolute improvement of 11.6 percentage points. A similar and even more pronounced trend is observed for Llama-3.1-70B-Instruct, where the ASR climbs from 68.9\% to 82.0\%, a gain of 13.1 points. Even against the more robust GPT-o4-mini, the ASR improves from 21.2\% to 26.1\%. This trend confirms that exploring multiple linguistic instantiations of a given strategy allows the attacker to overcome the stochasticity of generation and find a more potent attack vector. However, we also observe diminishing returns, as the performance gain from N=4 to N=8 is less pronounced than from N=2 to N=4, suggesting a trade-off between computational cost and marginal ASR improvement.

The Beam Search method demonstrates even more potent results, particularly against the strongest target model. By exploring combinations of strategies instead of just utilizing a single one, this method uncovers more complex and effective attack vectors. For the Llama models, Beam Search with a beam width ($W$) of 8 achieves the highest ASR, reaching 81.3\% on the 8B model and 84.5\% on the 70B model. The most striking result is on GPT-o4-mini, where Beam Search at $W=8$ reaches an ASR of 33.7\%. This represents a 12.5 percentage point improvement over the baseline and is significantly higher than the 26.1\% ASR achieved by Best-of-N at N=8. This highlights the superior ability of Beam Search to find novel attack paths through strategy combination. A direct comparison of the two methods reveals a critical insight: while Best-of-N is effective at exploring jailbreak prompts from one single strategy, Beam Search excels at finding synergistic combinations of strategies. This capability is especially crucial for jailbreaking more advanced, heavily aligned models like GPT-o4-mini.
\section{Conclusions}\label{conclusion}
In this paper, we propose a simple yet effective enhancement of the original AutoDAN-Turbo~\citep{liu2025autodanturbo} by introducing test-time scaling methods. Our experiments demonstrate that both Best-of-N and Beam Search scaling significantly improve jailbreak success rates across multiple state-of-the-art LLMs, with Beam Search proving particularly effective at creating synergistic strategy combinations to defeat highly robust models. This work underscores the value of dedicating inference-time compute to more thoroughly explore a learned strategy space. The primary limitation of this approach is the increased computational cost and latency associated with generating and evaluating multiple candidate prompts.

\bibliography{iclr2025_conference}

\begin{thebibliography}{18}
\providecommand{\natexlab}[1]{#1}
\providecommand{\url}[1]{\texttt{#1}}
\expandafter\ifx\csname urlstyle\endcsname\relax
  \providecommand{\doi}[1]{doi: #1}\else
  \providecommand{\doi}{doi: \begingroup \urlstyle{rm}\Url}\fi

\bibitem[Beetham et~al.(2025)Beetham, Chakraborty, Wang, Huang, Bedi, and Shah]{beetham2025liarleveraginginferencetime}
James Beetham, Souradip Chakraborty, Mengdi Wang, Furong Huang, Amrit~Singh Bedi, and Mubarak Shah.
\newblock Liar: Leveraging inference time alignment (best-of-n) to jailbreak llms in seconds, 2025.
\newblock URL \url{https://arxiv.org/abs/2412.05232}.

\bibitem[Chao et~al.(2023)Chao, Robey, Dobriban, Hassani, Pappas, and Wong]{chao2023jailbreaking}
Patrick Chao, Alexander Robey, Edgar Dobriban, Hamed Hassani, George~J. Pappas, and Eric Wong.
\newblock Jailbreaking black box large language models in twenty queries, 2023.

\bibitem[DeepSeek-AI et~al.(2025)DeepSeek-AI, Guo, Yang, Zhang, Song, Zhang, Xu, Zhu, Ma, Wang, Bi, Zhang, Yu, Wu, Wu, Gou, Shao, Li, Gao, Liu, Xue, Wang, Wu, Feng, Lu, Zhao, Deng, Zhang, Ruan, Dai, Chen, Ji, Li, Lin, Dai, Luo, Hao, Chen, Li, Zhang, Bao, Xu, Wang, Ding, Xin, Gao, Qu, Li, Guo, Li, Wang, Chen, Yuan, Qiu, Li, Cai, Ni, Liang, Chen, Dong, Hu, Gao, Guan, Huang, Yu, Wang, Zhang, Zhao, Wang, Zhang, Xu, Xia, Zhang, Zhang, Tang, Li, Wang, Li, Tian, Huang, Zhang, Wang, Chen, Du, Ge, Zhang, Pan, Wang, Chen, Jin, Chen, Lu, Zhou, Chen, Ye, Wang, Yu, Zhou, Pan, Li, Zhou, Wu, Ye, Yun, Pei, Sun, Wang, Zeng, Zhao, Liu, Liang, Gao, Yu, Zhang, Xiao, An, Liu, Wang, Chen, Nie, Cheng, Liu, Xie, Liu, Yang, Li, Su, Lin, Li, Jin, Shen, Chen, Sun, Wang, Song, Zhou, Wang, Shan, Li, Wang, Wei, Zhang, Xu, Li, Zhao, Sun, Wang, Yu, Zhang, Shi, Xiong, He, Piao, Wang, Tan, Ma, Liu, Guo, Ou, Wang, Gong, Zou, He, Xiong, Luo, You, Liu, Zhou, Zhu, Xu, Huang, Li, Zheng, Zhu, Ma, Tang, Zha, Yan, Ren, Ren, Sha, Fu, Xu, Xie, Zhang,
  Hao, Ma, Yan, Wu, Gu, Zhu, Liu, Li, Xie, Song, Pan, Huang, Xu, Zhang, and Zhang]{deepseekai2025deepseekr1incentivizingreasoningcapability}
DeepSeek-AI, Daya Guo, Dejian Yang, Haowei Zhang, Junxiao Song, Ruoyu Zhang, Runxin Xu, Qihao Zhu, Shirong Ma, Peiyi Wang, Xiao Bi, Xiaokang Zhang, Xingkai Yu, Yu~Wu, Z.~F. Wu, Zhibin Gou, Zhihong Shao, Zhuoshu Li, Ziyi Gao, Aixin Liu, Bing Xue, Bingxuan Wang, Bochao Wu, Bei Feng, Chengda Lu, Chenggang Zhao, Chengqi Deng, Chenyu Zhang, Chong Ruan, Damai Dai, Deli Chen, Dongjie Ji, Erhang Li, Fangyun Lin, Fucong Dai, Fuli Luo, Guangbo Hao, Guanting Chen, Guowei Li, H.~Zhang, Han Bao, Hanwei Xu, Haocheng Wang, Honghui Ding, Huajian Xin, Huazuo Gao, Hui Qu, Hui Li, Jianzhong Guo, Jiashi Li, Jiawei Wang, Jingchang Chen, Jingyang Yuan, Junjie Qiu, Junlong Li, J.~L. Cai, Jiaqi Ni, Jian Liang, Jin Chen, Kai Dong, Kai Hu, Kaige Gao, Kang Guan, Kexin Huang, Kuai Yu, Lean Wang, Lecong Zhang, Liang Zhao, Litong Wang, Liyue Zhang, Lei Xu, Leyi Xia, Mingchuan Zhang, Minghua Zhang, Minghui Tang, Meng Li, Miaojun Wang, Mingming Li, Ning Tian, Panpan Huang, Peng Zhang, Qiancheng Wang, Qinyu Chen, Qiushi Du, Ruiqi Ge, Ruisong
  Zhang, Ruizhe Pan, Runji Wang, R.~J. Chen, R.~L. Jin, Ruyi Chen, Shanghao Lu, Shangyan Zhou, Shanhuang Chen, Shengfeng Ye, Shiyu Wang, Shuiping Yu, Shunfeng Zhou, Shuting Pan, S.~S. Li, Shuang Zhou, Shaoqing Wu, Shengfeng Ye, Tao Yun, Tian Pei, Tianyu Sun, T.~Wang, Wangding Zeng, Wanjia Zhao, Wen Liu, Wenfeng Liang, Wenjun Gao, Wenqin Yu, Wentao Zhang, W.~L. Xiao, Wei An, Xiaodong Liu, Xiaohan Wang, Xiaokang Chen, Xiaotao Nie, Xin Cheng, Xin Liu, Xin Xie, Xingchao Liu, Xinyu Yang, Xinyuan Li, Xuecheng Su, Xuheng Lin, X.~Q. Li, Xiangyue Jin, Xiaojin Shen, Xiaosha Chen, Xiaowen Sun, Xiaoxiang Wang, Xinnan Song, Xinyi Zhou, Xianzu Wang, Xinxia Shan, Y.~K. Li, Y.~Q. Wang, Y.~X. Wei, Yang Zhang, Yanhong Xu, Yao Li, Yao Zhao, Yaofeng Sun, Yaohui Wang, Yi~Yu, Yichao Zhang, Yifan Shi, Yiliang Xiong, Ying He, Yishi Piao, Yisong Wang, Yixuan Tan, Yiyang Ma, Yiyuan Liu, Yongqiang Guo, Yuan Ou, Yuduan Wang, Yue Gong, Yuheng Zou, Yujia He, Yunfan Xiong, Yuxiang Luo, Yuxiang You, Yuxuan Liu, Yuyang Zhou, Y.~X. Zhu,
  Yanhong Xu, Yanping Huang, Yaohui Li, Yi~Zheng, Yuchen Zhu, Yunxian Ma, Ying Tang, Yukun Zha, Yuting Yan, Z.~Z. Ren, Zehui Ren, Zhangli Sha, Zhe Fu, Zhean Xu, Zhenda Xie, Zhengyan Zhang, Zhewen Hao, Zhicheng Ma, Zhigang Yan, Zhiyu Wu, Zihui Gu, Zijia Zhu, Zijun Liu, Zilin Li, Ziwei Xie, Ziyang Song, Zizheng Pan, Zhen Huang, Zhipeng Xu, Zhongyu Zhang, and Zhen Zhang.
\newblock Deepseek-r1: Incentivizing reasoning capability in llms via reinforcement learning, 2025.
\newblock URL \url{https://arxiv.org/abs/2501.12948}.

\bibitem[Dubey et~al.(2024)Dubey, Jauhri, Pandey, Kadian, Al-Dahle, Letman, Mathur, Schelten, Yang, Fan, Goyal, Hartshorn, Yang, Mitra, Sravankumar, Korenev, Hinsvark, Rao, Zhang, Rodriguez, Gregerson, Spataru, Roziere, Biron, Tang, Chern, Caucheteux, Nayak, Bi, Marra, McConnell, Keller, Touret, Wu, Wong, Ferrer, Nikolaidis, Allonsius, Song, Pintz, Livshits, Esiobu, Choudhary, Mahajan, Garcia-Olano, Perino, Hupkes, Lakomkin, AlBadawy, Lobanova, Dinan, Smith, Radenovic, Zhang, Synnaeve, Lee, Anderson, Nail, Mialon, Pang, Cucurell, Nguyen, Korevaar, Xu, Touvron, Zarov, Ibarra, Kloumann, Misra, Evtimov, Copet, Lee, Geffert, Vranes, Park, Mahadeokar, Shah, van~der Linde, Billock, Hong, Lee, Fu, Chi, Huang, Liu, Wang, Yu, Bitton, Spisak, Park, Rocca, Johnstun, Saxe, Jia, Alwala, Upasani, Plawiak, Li, Heafield, Stone, El-Arini, Iyer, Malik, Chiu, Bhalla, Rantala-Yeary, van~der Maaten, Chen, Tan, Jenkins, Martin, Madaan, Malo, Blecher, Landzaat, de~Oliveira, Muzzi, Pasupuleti, Singh, Paluri, Kardas, Oldham, Rita,
  Pavlova, Kambadur, Lewis, Si, Singh, Hassan, Goyal, Torabi, Bashlykov, Bogoychev, Chatterji, Duchenne, Çelebi, Alrassy, Zhang, Li, Vasic, Weng, Bhargava, Dubal, Krishnan, Koura, Xu, He, Dong, Srinivasan, Ganapathy, Calderer, Cabral, Stojnic, Raileanu, Girdhar, Patel, Sauvestre, Polidoro, Sumbaly, Taylor, Silva, Hou, Wang, Hosseini, Chennabasappa, Singh, Bell, Kim, Edunov, Nie, Narang, Raparthy, Shen, Wan, Bhosale, Zhang, Vandenhende, Batra, Whitman, Sootla, Collot, Gururangan, Borodinsky, Herman, Fowler, Sheasha, Georgiou, Scialom, Speckbacher, Mihaylov, Xiao, Karn, Goswami, Gupta, Ramanathan, Kerkez, Gonguet, Do, Vogeti, Petrovic, Chu, Xiong, Fu, Meers, Martinet, Wang, Tan, Xie, Jia, Wang, Goldschlag, Gaur, Babaei, Wen, Song, Zhang, Li, Mao, Coudert, Yan, Chen, Papakipos, Singh, Grattafiori, Jain, Kelsey, Shajnfeld, Gangidi, Victoria, Goldstand, Menon, Sharma, Boesenberg, Vaughan, Baevski, Feinstein, Kallet, Sangani, Yunus, Lupu, Alvarado, Caples, Gu, Ho, Poulton, Ryan, Ramchandani, Franco, Saraf,
  Chowdhury, Gabriel, Bharambe, Eisenman, Yazdan, James, Maurer, Leonhardi, Huang, Loyd, Paola, Paranjape, Liu, Wu, Ni, Hancock, Wasti, Spence, Stojkovic, Gamido, Montalvo, Parker, Burton, Mejia, Wang, Kim, Zhou, Hu, Chu, Cai, Tindal, Feichtenhofer, Civin, Beaty, Kreymer, Li, Wyatt, Adkins, Xu, Testuggine, David, Parikh, Liskovich, Foss, Wang, Le, Holland, Dowling, Jamil, Montgomery, Presani, Hahn, Wood, Brinkman, Arcaute, Dunbar, Smothers, Sun, Kreuk, Tian, Ozgenel, Caggioni, Guzmán, Kanayet, Seide, Florez, Schwarz, Badeer, Swee, Halpern, Thattai, Herman, Sizov, Guangyi, Zhang, Lakshminarayanan, Shojanazeri, Zou, Wang, Zha, Habeeb, Rudolph, Suk, Aspegren, Goldman, Damlaj, Molybog, Tufanov, Veliche, Gat, Weissman, Geboski, Kohli, Asher, Gaya, Marcus, Tang, Chan, Zhen, Reizenstein, Teboul, Zhong, Jin, Yang, Cummings, Carvill, Shepard, McPhie, Torres, Ginsburg, Wang, Wu, U, Saxena, Prasad, Khandelwal, Zand, Matosich, Veeraraghavan, Michelena, Li, Huang, Chawla, Lakhotia, Huang, Chen, Garg, A, Silva, Bell,
  Zhang, Guo, Yu, Moshkovich, Wehrstedt, Khabsa, Avalani, Bhatt, Tsimpoukelli, Mankus, Hasson, Lennie, Reso, Groshev, Naumov, Lathi, Keneally, Seltzer, Valko, Restrepo, Patel, Vyatskov, Samvelyan, Clark, Macey, Wang, Hermoso, Metanat, Rastegari, Bansal, Santhanam, Parks, White, Bawa, Singhal, Egebo, Usunier, Laptev, Dong, Zhang, Cheng, Chernoguz, Hart, Salpekar, Kalinli, Kent, Parekh, Saab, Balaji, Rittner, Bontrager, Roux, Dollar, Zvyagina, Ratanchandani, Yuvraj, Liang, Alao, Rodriguez, Ayub, Murthy, Nayani, Mitra, Li, Hogan, Battey, Wang, Maheswari, Howes, Rinott, Bondu, Datta, Chugh, Hunt, Dhillon, Sidorov, Pan, Verma, Yamamoto, Ramaswamy, Lindsay, Lindsay, Feng, Lin, Zha, Shankar, Zhang, Zhang, Wang, Agarwal, Sajuyigbe, Chintala, Max, Chen, Kehoe, Satterfield, Govindaprasad, Gupta, Cho, Virk, Subramanian, Choudhury, Goldman, Remez, Glaser, Best, Kohler, Robinson, Li, Zhang, Matthews, Chou, Shaked, Vontimitta, Ajayi, Montanez, Mohan, Kumar, Mangla, Albiero, Ionescu, Poenaru, Mihailescu, Ivanov, Li, Wang,
  Jiang, Bouaziz, Constable, Tang, Wang, Wu, Wang, Xia, Wu, Gao, Chen, Hu, Jia, Qi, Li, Zhang, Zhang, Adi, Nam, Yu, Wang, Hao, Qian, He, Rait, DeVito, Rosnbrick, Wen, Yang, and Zhao]{dubey2024llama3herdmodels}
Abhimanyu Dubey, Abhinav Jauhri, Abhinav Pandey, Abhishek Kadian, Ahmad Al-Dahle, Aiesha Letman, Akhil Mathur, Alan Schelten, Amy Yang, Angela Fan, Anirudh Goyal, Anthony Hartshorn, Aobo Yang, Archi Mitra, Archie Sravankumar, Artem Korenev, Arthur Hinsvark, Arun Rao, Aston Zhang, Aurelien Rodriguez, Austen Gregerson, Ava Spataru, Baptiste Roziere, Bethany Biron, Binh Tang, Bobbie Chern, Charlotte Caucheteux, Chaya Nayak, Chloe Bi, Chris Marra, Chris McConnell, Christian Keller, Christophe Touret, Chunyang Wu, Corinne Wong, Cristian~Canton Ferrer, Cyrus Nikolaidis, Damien Allonsius, Daniel Song, Danielle Pintz, Danny Livshits, David Esiobu, Dhruv Choudhary, Dhruv Mahajan, Diego Garcia-Olano, Diego Perino, Dieuwke Hupkes, Egor Lakomkin, Ehab AlBadawy, Elina Lobanova, Emily Dinan, Eric~Michael Smith, Filip Radenovic, Frank Zhang, Gabriel Synnaeve, Gabrielle Lee, Georgia~Lewis Anderson, Graeme Nail, Gregoire Mialon, Guan Pang, Guillem Cucurell, Hailey Nguyen, Hannah Korevaar, Hu~Xu, Hugo Touvron, Iliyan Zarov,
  Imanol~Arrieta Ibarra, Isabel Kloumann, Ishan Misra, Ivan Evtimov, Jade Copet, Jaewon Lee, Jan Geffert, Jana Vranes, Jason Park, Jay Mahadeokar, Jeet Shah, Jelmer van~der Linde, Jennifer Billock, Jenny Hong, Jenya Lee, Jeremy Fu, Jianfeng Chi, Jianyu Huang, Jiawen Liu, Jie Wang, Jiecao Yu, Joanna Bitton, Joe Spisak, Jongsoo Park, Joseph Rocca, Joshua Johnstun, Joshua Saxe, Junteng Jia, Kalyan~Vasuden Alwala, Kartikeya Upasani, Kate Plawiak, Ke~Li, Kenneth Heafield, Kevin Stone, Khalid El-Arini, Krithika Iyer, Kshitiz Malik, Kuenley Chiu, Kunal Bhalla, Lauren Rantala-Yeary, Laurens van~der Maaten, Lawrence Chen, Liang Tan, Liz Jenkins, Louis Martin, Lovish Madaan, Lubo Malo, Lukas Blecher, Lukas Landzaat, Luke de~Oliveira, Madeline Muzzi, Mahesh Pasupuleti, Mannat Singh, Manohar Paluri, Marcin Kardas, Mathew Oldham, Mathieu Rita, Maya Pavlova, Melanie Kambadur, Mike Lewis, Min Si, Mitesh~Kumar Singh, Mona Hassan, Naman Goyal, Narjes Torabi, Nikolay Bashlykov, Nikolay Bogoychev, Niladri Chatterji, Olivier
  Duchenne, Onur Çelebi, Patrick Alrassy, Pengchuan Zhang, Pengwei Li, Petar Vasic, Peter Weng, Prajjwal Bhargava, Pratik Dubal, Praveen Krishnan, Punit~Singh Koura, Puxin Xu, Qing He, Qingxiao Dong, Ragavan Srinivasan, Raj Ganapathy, Ramon Calderer, Ricardo~Silveira Cabral, Robert Stojnic, Roberta Raileanu, Rohit Girdhar, Rohit Patel, Romain Sauvestre, Ronnie Polidoro, Roshan Sumbaly, Ross Taylor, Ruan Silva, Rui Hou, Rui Wang, Saghar Hosseini, Sahana Chennabasappa, Sanjay Singh, Sean Bell, Seohyun~Sonia Kim, Sergey Edunov, Shaoliang Nie, Sharan Narang, Sharath Raparthy, Sheng Shen, Shengye Wan, Shruti Bhosale, Shun Zhang, Simon Vandenhende, Soumya Batra, Spencer Whitman, Sten Sootla, Stephane Collot, Suchin Gururangan, Sydney Borodinsky, Tamar Herman, Tara Fowler, Tarek Sheasha, Thomas Georgiou, Thomas Scialom, Tobias Speckbacher, Todor Mihaylov, Tong Xiao, Ujjwal Karn, Vedanuj Goswami, Vibhor Gupta, Vignesh Ramanathan, Viktor Kerkez, Vincent Gonguet, Virginie Do, Vish Vogeti, Vladan Petrovic, Weiwei Chu,
  Wenhan Xiong, Wenyin Fu, Whitney Meers, Xavier Martinet, Xiaodong Wang, Xiaoqing~Ellen Tan, Xinfeng Xie, Xuchao Jia, Xuewei Wang, Yaelle Goldschlag, Yashesh Gaur, Yasmine Babaei, Yi~Wen, Yiwen Song, Yuchen Zhang, Yue Li, Yuning Mao, Zacharie~Delpierre Coudert, Zheng Yan, Zhengxing Chen, Zoe Papakipos, Aaditya Singh, Aaron Grattafiori, Abha Jain, Adam Kelsey, Adam Shajnfeld, Adithya Gangidi, Adolfo Victoria, Ahuva Goldstand, Ajay Menon, Ajay Sharma, Alex Boesenberg, Alex Vaughan, Alexei Baevski, Allie Feinstein, Amanda Kallet, Amit Sangani, Anam Yunus, Andrei Lupu, Andres Alvarado, Andrew Caples, Andrew Gu, Andrew Ho, Andrew Poulton, Andrew Ryan, Ankit Ramchandani, Annie Franco, Aparajita Saraf, Arkabandhu Chowdhury, Ashley Gabriel, Ashwin Bharambe, Assaf Eisenman, Azadeh Yazdan, Beau James, Ben Maurer, Benjamin Leonhardi, Bernie Huang, Beth Loyd, Beto~De Paola, Bhargavi Paranjape, Bing Liu, Bo~Wu, Boyu Ni, Braden Hancock, Bram Wasti, Brandon Spence, Brani Stojkovic, Brian Gamido, Britt Montalvo, Carl
  Parker, Carly Burton, Catalina Mejia, Changhan Wang, Changkyu Kim, Chao Zhou, Chester Hu, Ching-Hsiang Chu, Chris Cai, Chris Tindal, Christoph Feichtenhofer, Damon Civin, Dana Beaty, Daniel Kreymer, Daniel Li, Danny Wyatt, David Adkins, David Xu, Davide Testuggine, Delia David, Devi Parikh, Diana Liskovich, Didem Foss, Dingkang Wang, Duc Le, Dustin Holland, Edward Dowling, Eissa Jamil, Elaine Montgomery, Eleonora Presani, Emily Hahn, Emily Wood, Erik Brinkman, Esteban Arcaute, Evan Dunbar, Evan Smothers, Fei Sun, Felix Kreuk, Feng Tian, Firat Ozgenel, Francesco Caggioni, Francisco Guzmán, Frank Kanayet, Frank Seide, Gabriela~Medina Florez, Gabriella Schwarz, Gada Badeer, Georgia Swee, Gil Halpern, Govind Thattai, Grant Herman, Grigory Sizov, Guangyi, Zhang, Guna Lakshminarayanan, Hamid Shojanazeri, Han Zou, Hannah Wang, Hanwen Zha, Haroun Habeeb, Harrison Rudolph, Helen Suk, Henry Aspegren, Hunter Goldman, Ibrahim Damlaj, Igor Molybog, Igor Tufanov, Irina-Elena Veliche, Itai Gat, Jake Weissman, James
  Geboski, James Kohli, Japhet Asher, Jean-Baptiste Gaya, Jeff Marcus, Jeff Tang, Jennifer Chan, Jenny Zhen, Jeremy Reizenstein, Jeremy Teboul, Jessica Zhong, Jian Jin, Jingyi Yang, Joe Cummings, Jon Carvill, Jon Shepard, Jonathan McPhie, Jonathan Torres, Josh Ginsburg, Junjie Wang, Kai Wu, Kam~Hou U, Karan Saxena, Karthik Prasad, Kartikay Khandelwal, Katayoun Zand, Kathy Matosich, Kaushik Veeraraghavan, Kelly Michelena, Keqian Li, Kun Huang, Kunal Chawla, Kushal Lakhotia, Kyle Huang, Lailin Chen, Lakshya Garg, Lavender A, Leandro Silva, Lee Bell, Lei Zhang, Liangpeng Guo, Licheng Yu, Liron Moshkovich, Luca Wehrstedt, Madian Khabsa, Manav Avalani, Manish Bhatt, Maria Tsimpoukelli, Martynas Mankus, Matan Hasson, Matthew Lennie, Matthias Reso, Maxim Groshev, Maxim Naumov, Maya Lathi, Meghan Keneally, Michael~L. Seltzer, Michal Valko, Michelle Restrepo, Mihir Patel, Mik Vyatskov, Mikayel Samvelyan, Mike Clark, Mike Macey, Mike Wang, Miquel~Jubert Hermoso, Mo~Metanat, Mohammad Rastegari, Munish Bansal, Nandhini
  Santhanam, Natascha Parks, Natasha White, Navyata Bawa, Nayan Singhal, Nick Egebo, Nicolas Usunier, Nikolay~Pavlovich Laptev, Ning Dong, Ning Zhang, Norman Cheng, Oleg Chernoguz, Olivia Hart, Omkar Salpekar, Ozlem Kalinli, Parkin Kent, Parth Parekh, Paul Saab, Pavan Balaji, Pedro Rittner, Philip Bontrager, Pierre Roux, Piotr Dollar, Polina Zvyagina, Prashant Ratanchandani, Pritish Yuvraj, Qian Liang, Rachad Alao, Rachel Rodriguez, Rafi Ayub, Raghotham Murthy, Raghu Nayani, Rahul Mitra, Raymond Li, Rebekkah Hogan, Robin Battey, Rocky Wang, Rohan Maheswari, Russ Howes, Ruty Rinott, Sai~Jayesh Bondu, Samyak Datta, Sara Chugh, Sara Hunt, Sargun Dhillon, Sasha Sidorov, Satadru Pan, Saurabh Verma, Seiji Yamamoto, Sharadh Ramaswamy, Shaun Lindsay, Shaun Lindsay, Sheng Feng, Shenghao Lin, Shengxin~Cindy Zha, Shiva Shankar, Shuqiang Zhang, Shuqiang Zhang, Sinong Wang, Sneha Agarwal, Soji Sajuyigbe, Soumith Chintala, Stephanie Max, Stephen Chen, Steve Kehoe, Steve Satterfield, Sudarshan Govindaprasad, Sumit Gupta,
  Sungmin Cho, Sunny Virk, Suraj Subramanian, Sy~Choudhury, Sydney Goldman, Tal Remez, Tamar Glaser, Tamara Best, Thilo Kohler, Thomas Robinson, Tianhe Li, Tianjun Zhang, Tim Matthews, Timothy Chou, Tzook Shaked, Varun Vontimitta, Victoria Ajayi, Victoria Montanez, Vijai Mohan, Vinay~Satish Kumar, Vishal Mangla, Vítor Albiero, Vlad Ionescu, Vlad Poenaru, Vlad~Tiberiu Mihailescu, Vladimir Ivanov, Wei Li, Wenchen Wang, Wenwen Jiang, Wes Bouaziz, Will Constable, Xiaocheng Tang, Xiaofang Wang, Xiaojian Wu, Xiaolan Wang, Xide Xia, Xilun Wu, Xinbo Gao, Yanjun Chen, Ye~Hu, Ye~Jia, Ye~Qi, Yenda Li, Yilin Zhang, Ying Zhang, Yossi Adi, Youngjin Nam, Yu, Wang, Yuchen Hao, Yundi Qian, Yuzi He, Zach Rait, Zachary DeVito, Zef Rosnbrick, Zhaoduo Wen, Zhenyu Yang, and Zhiwei Zhao.
\newblock The llama 3 herd of models, 2024.
\newblock URL \url{https://arxiv.org/abs/2407.21783}.

\bibitem[Jiang et~al.(2024)Jiang, Xu, Niu, Xiang, Ramasubramanian, Li, and Poovendran]{jiang2024artprompt}
Fengqing Jiang, Zhangchen Xu, Luyao Niu, Zhen Xiang, Bhaskar Ramasubramanian, Bo~Li, and Radha Poovendran.
\newblock Artprompt: Ascii art-based jailbreak attacks against aligned llms, 2024.

\bibitem[Liu et~al.(2024)Liu, Xu, Chen, and Xiao]{liu2024autodan}
Xiaogeng Liu, Nan Xu, Muhao Chen, and Chaowei Xiao.
\newblock Autodan: Generating stealthy jailbreak prompts on aligned large language models.
\newblock In \emph{The Twelfth International Conference on Learning Representations}, 2024.
\newblock URL \url{https://openreview.net/forum?id=7Jwpw4qKkb}.

\bibitem[Liu et~al.(2025)Liu, Li, Suh, Vorobeychik, Mao, Jha, McDaniel, Sun, Li, and Xiao]{liu2025autodanturbo}
Xiaogeng Liu, Peiran Li, G.~Edward Suh, Yevgeniy Vorobeychik, Zhuoqing Mao, Somesh Jha, Patrick McDaniel, Huan Sun, Bo~Li, and Chaowei Xiao.
\newblock Auto{DAN}-turbo: A lifelong agent for strategy self-exploration to jailbreak {LLM}s.
\newblock In \emph{The Thirteenth International Conference on Learning Representations}, 2025.
\newblock URL \url{https://openreview.net/forum?id=bhK7U37VW8}.

\bibitem[Lv et~al.(2024)Lv, Wang, Zhang, Huang, Dou, Ye, Gui, Zhang, and Huang]{lv2024codechameleon}
Huijie Lv, Xiao Wang, Yuansen Zhang, Caishuang Huang, Shihan Dou, Junjie Ye, Tao Gui, Qi~Zhang, and Xuanjing Huang.
\newblock Codechameleon: Personalized encryption framework for jailbreaking large language models, 2024.

\bibitem[Mazeika et~al.(2024)Mazeika, Phan, Yin, Zou, Wang, Mu, Sakhaee, Li, Basart, Li, Forsyth, and Hendrycks]{mazeika2024harmbench}
Mantas Mazeika, Long Phan, Xuwang Yin, Andy Zou, Zifan Wang, Norman Mu, Elham Sakhaee, Nathaniel Li, Steven Basart, Bo~Li, David Forsyth, and Dan Hendrycks.
\newblock Harmbench: A standardized evaluation framework for automated red teaming and robust refusal.
\newblock 2024.

\bibitem[OpenAI(2025)]{openai2025o3o4mini}
OpenAI.
\newblock Introducing o3 and o4-mini.
\newblock Web Page, April 2025.
\newblock URL \url{https://openai.com/index/introducing-o3-and-o4-mini/}.

\bibitem[Sabbaghi et~al.(2025)Sabbaghi, Kassianik, Pappas, Singer, Karbasi, and Hassani]{sabbaghi2025adversarialreasoningjailbreakingtime}
Mahdi Sabbaghi, Paul Kassianik, George Pappas, Yaron Singer, Amin Karbasi, and Hamed Hassani.
\newblock Adversarial reasoning at jailbreaking time, 2025.
\newblock URL \url{https://arxiv.org/abs/2502.01633}.

\bibitem[Shen et~al.(2023)Shen, Chen, Backes, Shen, and Zhang]{shen2023do}
Xinyue Shen, Zeyuan Chen, Michael Backes, Yun Shen, and Yang Zhang.
\newblock "do anything now": Characterizing and evaluating in-the-wild jailbreak prompts on large language models, 2023.

\bibitem[Snell et~al.(2024)Snell, Lee, Xu, and Kumar]{snell2024scalingllmtesttimecompute}
Charlie Snell, Jaehoon Lee, Kelvin Xu, and Aviral Kumar.
\newblock Scaling llm test-time compute optimally can be more effective than scaling model parameters, 2024.
\newblock URL \url{https://arxiv.org/abs/2408.03314}.

\bibitem[walkerspider(2022)]{walkerspider2023Dan}
walkerspider.
\newblock \url{https://old.reddit.com/r/ChatGPT/comments/zl cyr9/dan_is_my_new_friend/}, 2022.
\newblock Accessed: 2023-09-28.

\bibitem[Wei et~al.(2023)Wei, Haghtalab, and Steinhardt]{NEURIPS2023_fd661313}
Alexander Wei, Nika Haghtalab, and Jacob Steinhardt.
\newblock Jailbroken: How does llm safety training fail?
\newblock In \emph{Advances in Neural Information Processing Systems}, volume~36, pp.\  80079--80110, 2023.
\newblock URL \url{https://proceedings.neurips.cc/paper_files/paper/2023/file/fd6613131889a4b656206c50a8bd7790-Paper-Conference.pdf}.

\bibitem[Yuan et~al.(2024)Yuan, Jiao, Wang, tse Huang, He, Shi, and Tu]{yuan2024gpt4}
Youliang Yuan, Wenxiang Jiao, Wenxuan Wang, Jen tse Huang, Pinjia He, Shuming Shi, and Zhaopeng Tu.
\newblock Gpt-4 is too smart to be safe: Stealthy chat with llms via cipher, 2024.

\bibitem[Zeng et~al.(2024)Zeng, Lin, Zhang, Yang, Jia, and Shi]{zeng2024johnny}
Yi~Zeng, Hongpeng Lin, Jingwen Zhang, Diyi Yang, Ruoxi Jia, and Weiyan Shi.
\newblock How johnny can persuade llms to jailbreak them: Rethinking persuasion to challenge ai safety by humanizing llms, 2024.

\bibitem[Zou et~al.(2023)Zou, Wang, Carlini, Nasr, Kolter, and Fredrikson]{zou2023universal}
Andy Zou, Zifan Wang, Nicholas Carlini, Milad Nasr, J.~Zico Kolter, and Matt Fredrikson.
\newblock Universal and transferable adversarial attacks on aligned language models, 2023.

\end{thebibliography}
\bibliographystyle{iclr2025_conference}

\end{document}